# Patient-Centric Cellular Networks Optimization using Big Data Analytics

Mohammed S. Hadi, Ahmed Q. Lawey, Taisir E. H. El-Gorashi and J. M. H Elmirghani

*Abstract*— Big data analytics is one of the state-of-the-art tools to optimize networks and transform them from merely being a blind tube that conveys data, into a cognitive, conscious, and self-optimizing entity that can intelligently adapt according to the needs of its users. This, in fact, can be regarded as one of the highest forthcoming priorities of future networks. In this paper, we propose a system for Out-Patient (OP) centric Long Term Evolution-Advanced (LTE-A) network optimization. Big data harvested from the OPs' medical records, along with current readings from their body sensors are processed and analyzed to predict the likelihood of a life-threatening medical condition, for instance, an imminent stroke. This prediction is used to ensure that the OP is assigned an optimal LTE-A Physical Resource Blocks (PRBs) to transmit their critical data to their healthcare provider with minimal delay. To the best of our knowledge, this is the first time big data analytics are utilized to optimize a cellular network in an OP-conscious manner. The PRBs assignment is optimized using Mixed Integer Linear Programming (MILP) and a real-time heuristic. Two approaches are proposed, the Weighted Sum Rate Maximization (WSRMax) approach and the Proportional Fairness (PF) approach. The approaches increased the OPs' average SINR by 26.6% and 40.5%, respectively. The WSRMax approach increased the system's total SINR to a level higher than that of the PF approach, however, the PF approach reported higher SINRs for the OPs, better fairness and a lower margin of error.

## I. INTRODUCTION

PRIOR to the emergence of big data, decisions were made relying on data samples. Consequently, the decisions were semi-optimum [1]. Those ill-informed decisions spanned over different areas from marketing to law enforcement, sports, and healthcare. With the proliferation of social media applications, Internet of Things (IoT) sensors, and Global Positioning System (GPS)-based services, people may now be considered as walking generators of data. The powerful capability of big data analytics in analyzing massive amounts of data and inferring knowledge from it [2] has brought about better predictions paving the way for better decisions.

Healthcare is a vital subject due to its role in people's lives. The continuous increase in the world population and other factors, like insufficient healthcare budgets, has resulted in crowded hospitals, over-worked medical staff, and extended queuing times for the patients. Given the global nature of the problem, researchers are developing new approaches to improve the level of care delivered by healthcare providers while ensuring a reduction in all previously-mentioned points. Big data can be used to ensure medical service is reaching those most in need, in a timely manner [3]. Big data analytics can provide accurate diagnosis by offering the ability to analyze and infer from the patient's history, their daily routine, diet, allergies, and genetic information, etc. Such analyses can be time consuming and requires certain level of expertise to be carried out by medical personnel [4]. An example mentioned in [5] reports the use of big data analytics by Columbia University Medical Centre to diagnose complications in patients with bleeding stroke caused by raptured brain aneurysm. Based on physiological data, the diagnosis was reported 48 hours beforehand in patients with brain injuries, which gave the medical professionals a head start to address these complications.

In the healthcare sector, there are many sources of big data, for example; IoT medically-related sensors, smart watches, and smartphone medical applications. What the above-mentioned data generators have in common is their reliance on network connectivity. Maintaining this connectivity and ensuring its quality is a dilemma that many researchers tried to solve optimally. Here, the patient's big data can play a double role. In addition to diagnosis, it can guide the network operator to the patients who have the highest and most urgent needs, and thus direct their network resources towards these patients. We believe that ensuring high quality connectivity between the patient-linked peripherals and their healthcare provider is an important step towards highly personalized e-healthcare services and applications.

A wireless connection is preferred over a wired one for what it has to offer in terms of mobility. Consequently, cellular and Wi-Fi are the most popular connectivity technologies. The level of freedom (mobility-wise) varies between wireless technologies, for example, Wi-Fi may provide an adequate data rate, nevertheless, it forces an Out-Patient (OP) that needs to keep his/her medical IoT sensor (e.g. IoT pacemaker) connected, to stay within a relatively-small coverage area (i.e., indoors mainly). Utilizing the already-existing cellular networks can provide a much-needed freedom to that OP. However, due to path loss and fading, this approach faces several problems because there might be some blind-spots, deeply-faded locations, where the Signal to Interference plus Noise Ratio (SINR) level is so low that the connection is unreliable or cannot be established. In a slow fading channel, this could mean that the signal level may not be adequate at the instant(s) when critical information relating to the OP's health has to be conveyed *immediately* to the health care provider. Big data is portrayed in [6] as a next generation tool that can be used to find an optimal trade-off problem between resource sharing, allocation, and optimization in wireless networks. Nevertheless, optimizing cellular networks in a user-centric style is still underexplored. In this paper, we introduce for the first time two OP-conscious approaches optimizing the uplink side of a multi-cell Orthogonal Frequency Division Multiple Access (OFDMA) network. In both models, the objective function prioritizes the OPs by maximizing their SINR received at the Base Station (BS) while keeping the goal of maximizing the network's overall SINR.

The models comprise an assignment scheme powered by big data analytics where OPs are assigned Physical Resource Blocks (PRBs) with powers proportional to their current medical situation. Fairness was incorporated to minimize the negative impact of such assignment on other users. The models are subject to several power and PRB assignment constrains that govern its operation. The main contributions of this paper are: (i) the use of big data analytics to prioritize cellular-connected-OPs and grant them suitable PRBs according to their health condition. (ii) The development of a mathematical method to determine the likelihood of a stroke by using a naïve Bayesian classifier and real patient big data sets. (iii) The introduction of an interdisciplinary approach to optimize the uplink of a Long Term Evolution-Advanced (LTE-A) network using big data analytics and MILP optimization.

The remainder of this paper is organized as follows. Section II discusses the related work. Section III presents the proposed system and the MILP formulation of the PRB assignment optimization problem. A real-time heuristic for PRBs assignment is presented in Section IV. Section V presents and discusses the results. The open research challenges are highlighted in section VI. Section VIII concludes the paper.

## II. RELATED WORK

Due to the nature of our proposed system, there are fundamentally two parts that need to be investigated in this section. The first part is concerned with the use of big data analytics for resource allocation and optimization in a cellular network. The second part focuses on the use of big data analytics to support the healthcare sector. This section concludes with a third part illustrating the link that we are proposing between the former two parts to forge a cellular network optimized to serve outpatients by reacting according to their needs.

### A. *Using Big Data Analytics for Cellular Networks Resource Allocation*

The topic of utilizing big data analytics in network design was thoroughly discussed in our survey paper. We observed that the highest number of papers in this area are in the wireless field [2]. Significant effort is dedicated currently to endowing wireless cellular networks with the ability to seamlessly *prioritize* users and serve them accordingly. Previous work in this area includes the work of the authors in [6] who proposed the use of configuration, alarm, and log files and processing the mentioned data using a big data processing environment, thus identifying the behavior of both the user and the network. The goal is to solve the problem of radio resource allocation to users in the Radio Access Network (RAN) in a manner that ensures minimal delay between resource request and assignment. Another idea was presented by the authors of [7] to manage the network resources in Heterogeneous Networks (HetNets). This was achieved through the utilization of sentimental and behavioral analysis of data collected from social networks, along with communication network data. The latter was exploited to predict sudden increases in usage of the mobile network. The aim was to achieve minimal service disruption by servicing the right place at the right time.

### B. *Using Big Data Analytics in Healthcare*

Several approaches have attempted to address the riddle of employing big data analytics to accomplish the task of OP monitoring. A system that has a real time response when an emergency case arises was proposed by the authors in [8]. The system is capable of processing data collected from millions of Wireless Body Area Network (WBAN) sensors. The authors of [9] investigated the challenges associated with designing and implementing big data services that utilize data harvested from medical sensors as well as other IoT applications. They also considered the requirement of processing this data in real-time. Another approach to help patients with Parkinson's disease was proposed by the authors of [10]. The system monitors the loss of flexibility as it is a sign of disease progression. This is done by analyzing big data collected from the body and 3D sensors, such as the Microsoft Kinect sensor system. The disease development and treatment effectiveness can both be observed by the patients as well as their healthcare providers in real-time. A survey conducted by the authors in [11] showed different approaches to detect heart disease at an early stage. The common theme among those approaches is that they are all based on data mining, machine learning, and big data analytics techniques.

### C. *Missing Piece of the Jigsaw*

All the approaches mentioned in the previous subsection assumed networks with ideal connectivity. However, in a real world scenario, opposing elements like channel fading and noise need to be taken into consideration. Our approach exploits big data analytics for the purpose of optimizing the Radio Access Network (RAN) side of an LTE-A network to serve a specific category of people, in this case, the OPs. Our approach ensures service availability to OPs, especially at times when they are in desperate need for it. We argue that by analyzing the OPs' big data we can predict the ones that are at high risk of having a stroke. Consequently, OPs will be prioritized over normal users and the network's attention (in terms of the quality of the assigned resources) can be shifted towards them. In the US, about 795 thousand people suffer a stroke annually [12]. This is equivalent to 1.5 stroke incidents per minute on average which is significant and frequent. In England, Northern Ireland and Wales, one third of stroke patients went to the hospital during 2016-2017 not knowing what time their symptoms commenced [13]. The problem is serious given an average time from the start of the symptoms till admission to a hospital of 7.5 hours, with another 55 minutes door-to-needle time (duration between arrival at the emergency department and administering an anesthetic) and the fact that a stroke patient is losing 1.9 million neurons each minute before treatment commence [13]. The use of our proposed system can have a tremendous impact on minimizing this time since patients are prioritized and given reliable resource. Moreover, the increase in the SINR will result in an increase in the spectral efficiency hence fewer resources are required to transmit the same amount of data [14]. The proposed system can also help in providing reliable connectivity to medical IoT devices when transmitting the patient's vital signs to the healthcare provider. In addition, it can help with early detection of symptoms and facilitate early emergency admittance to the hospital to help save patients' lives. If other forms of ill health are included, the proposed system will be called upon even more frequently.

## III. OP-CENTRIC NETWORK OPTIMIZATION MODEL

In this section, we present the system model, then we describe the problem formulation. For that purpose, a set of mathematical programming formulations adopted throughout this paper is presented.

### A. System Model

We consider an urban environment covered by an LTE-A cellular network. The area is populated with a number of users scattered at random distances from the BSs (between 300 and 600 meters). The users fall into two categories; normal (healthy) users and OPs as shown in Fig.1. As we previously indicated, cellular networks can provide an optimal way for OPs to have a connection. Since OPs are randomly-located, different power levels (signal strengths) will be received from their mobile devices. OPs with a higher likelihood of stroke must transmit their data as soon as possible. However, if the OP was assigned a channel with a low SINR, the required medical response may not arrive in time.

The goal is to prioritize OPs over normal users in terms of resource allocation.

The OP data is analyzed in a cloud-located big data analytics engine running a naïve Bayesian classifier, one of big data analytics algorithms [15]. This engine is used to predict the stroke likelihood for an OP. Based on this likelihood, the OPs are assigned proportional weights (i.e. priorities) to grant them PRBs with an optimal SINR favoring them over normal (i.e., healthy) users. Towards this end, the objective function of our optimization model guarantees the allocation of high gain PRBs to OPs, aiming at maximizing the total SINR received at the BS, and preserves fairness among users to ensure such a resource allocation scheme will not negatively-impact other users. We note that the terms 'healthy user' and 'normal user' are used interchangeably throughout this paper.

### B. Naïve Bayesian Classifier

We used the naïve Bayesian classifier to determine the likelihood of occurrence of a certain incident $c$ (e.g., a stroke) relying on a given set of independent feature variables $f_i$ obtained from the OPs' big data (i.e. medical records). Given, a *current state* of a certain OP, the classifier can use the training dataset (medical record) to determine the likelihood that this OP would suffer a stroke and quantify it as a risk factor. These feature variables represent the vital readings (e.g., Systolic and Diastolic blood pressure, total cholesterol, and smoking rate) that can be collected by body-attached IoT sensors and fed to the big data analytics engine where the naïve Bayesian classifier resides. It is worth noting that this classifier is termed *naïve* due to the assumption that the feature variables are conditionally independent [16]. We chose to use the naïve Bayesian classifier here for several reasons;

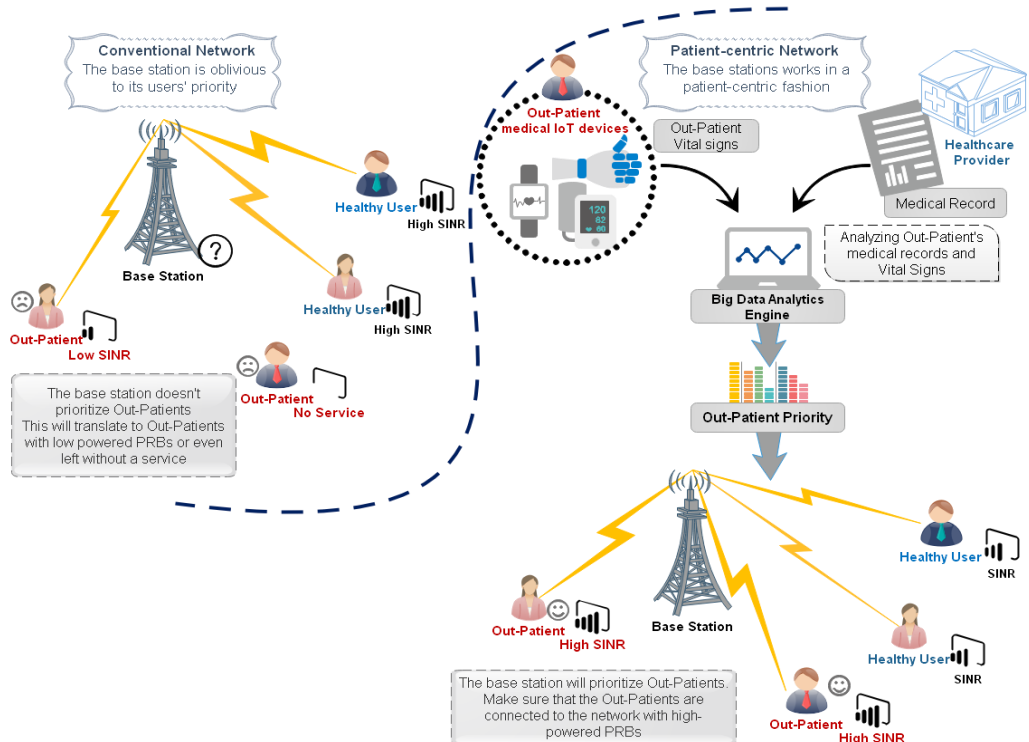

Fig. 1. Patient-Centric Cellular Network

(i) it proved to be one of the most practical approaches in solving learning problems, (ii) its confirmed competitiveness when compared to other algorithms including neural networks and decision trees [16], (iii) its low complexity and computation burden compared to other classifiers, and (iv) because it can be programmed jointly with the MILP and the heuristic used to optimize the uplink of the LTE-A network (as shown in Sections III and IV). Moreover, its effectiveness was proven in cardiovascular disease risk discovery as it was employed by the authors of [17] and had the highest accuracy among other approaches [18, 19].

The *likelihood* of $F$ given $C$ is given as

$$p(F_i = f_i | C = c) = \frac{\sum_{i=1}^{n}(C = c \wedge F_i = f_i)}{\sum_{i=1}^{n}(C_i = C_i)} \quad (1)$$

The naïve Bayesian classifier's *posterior probability* can be expressed as shown in equation (2).

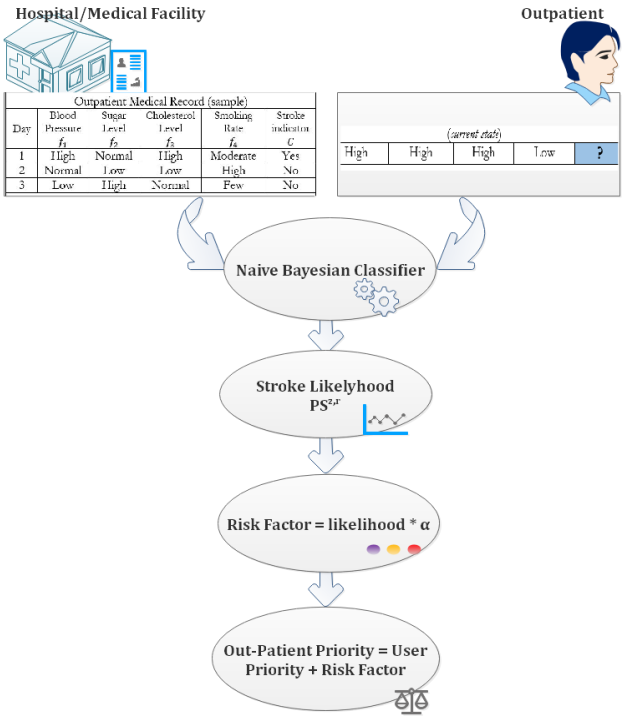

**Hospital/Medical Facility**

**Outpatient**

| | Outpatient Medical Record (sample) | | | | |
|---|---|---|---|---|---|
| Day | Blood Pressure Level $f_1$ | Sugar Level $f_2$ | Cholesterol Level $f_3$ | Smoking Rate $f_4$ | Stroke indicator $C$ |
| 1 | High | Normal | High | Moderate | Yes |
| 2 | Normal | Low | Low | High | No |
| 3 | Low | High | Normal | Few | No |

(current state)

| High | High | High | Low | ? |

**Naïve Bayesian Classifier**

**Stroke Likelihood** $PS^{t,t}$

**Risk Factor = likelihood * α**

**Out-Patient Priority = User Priority + Risk Factor**

Fig. 2. Naïve Bayesian Classifier Role / User Weight Calculation Procedure

$$p(C = c | F_i = f_i) = P(C = c) \prod_{i=1}^{n} P(Fi = fi | C = c) \quad (2)$$

where $P(C = c)$ represents *the prior probability* of stroke, in other words it is the number of days in which a stroke occurred over the total number of days (i.e. observation period). While $\prod_{i=1}^{n} P(Fi = fi | C = c)$ represents the *joint probability*.

The monitored body readings represented by feature variables $f_1, \dots f_4$ can take three levels reflecting the recorded state of each feature at a specific time of the day as shown in Table I. The class variable $C$ registers whether a stroke occurred in the corresponding day.

The role of the naïve Bayesian classifier is illustrated in Fig. 2. The classifier reads the OP's *medical record* (check table I) and uses the OP's *current state* (the lower part in Table I) to predict the likelihood of an upcoming stroke. This likelihood is to be converted later (in the upcoming subsection) into a risk factor used to calculate the weight given to each OP to be prioritized among other users during PRB assignment which is implemented in this work using a MILP and a heuristic, as explained in the subsequent subsection C. We also note that the terms "user weight" and "user priority" are used interchangeably throughout this paper.

Since preserving the patient's privacy is of utmost importance for healthcare providers, acquiring the above-mentioned numerical data values is not easy. We are fortunate in that the Framingham heart study in [20] has a big data set that covers the features we needed. We therefore populated the OP datasets by relying on this bid data, data set. The data in [20] contains readings from over three thousand patients, part of it was segmented to represent several OPs. Thus, each OP has a complete medical dataset. This cardiovascular cohort study that started in 1948 targeted adults residing in the town of Framingham, Massachusetts. The study is ongoing, and a new phase has started in 2002 with the enrollment of the third generation of participants [21]. The above–mentioned OP data has the characteristics of big data; hence, big data analytics algorithms can be used to predict the likelihood of occurrence of a certain incident (i.e. a stroke in our case).

TABLE I

OUTPATIENT MEDICAL RECORD (SAMPLE)

| Day | Blood Pressure $f_1$ | Sugar Level $f_2$ | Cholesterol Level $f_3$ | Smoking Rate $f_4$ | Stroke indicator $C$ |
|---|---|---|---|---|---|
| 1 | High | Normal | High | Moderate | Yes |
| 2 | Normal | Low | Low | High | No |
| 3 | Low | High | Normal | Few | No |
| | | (CURRENT STATE) | | | |
| High | High | High | Low | | ? |

For the Framingham dataset to be efficiently utilized by the naïve Bayesian classifier, it had to undergo several *stages of data preprocessing*. These stages involve:

1- Data Reduction

In this process, particular features are retained while others excluded. According to [22, 23] Hyperlipidemia (i.e. Total Cholesterol), blood pressure, and smoking are among the main contributors to a stroke. Hence, the selected features in this paper.

2- Data Cleansing

Incomplete, erroneous, and inconsistent entries were omitted. Thus, the resulting dataset is error-free and have a complete set of values across all entries.

3- Data Generalization

We categorized the feature values mentioned in Table. I according to their medical range shown in Table II. Using [24], the corresponding values where categorized into three severity levels for modelling purposes as shown in Table II, except for the smoking rate, which we categorized into the levels: light, moderate, and heavy, respectively as proposed in [25].

TABLE II
FEATURE VALUES AND THEIR CORRESPONDING LEVEL

| Feature | Range | Level |
|---|---|---|
| **Total cholesterol Level (mg/dl) [24]** | <200 | Optimal |
| | 200-239 | Normal |
| | 240+ | High |
| **Systolic BP (mmHg ) [24]** | <120 | Normal |
| | 120-139 | Pre-hypertension |
| | 140+ | High Hypertension |
| **Diastolic BP (mmHg) [24]** | <80 | Normal |
| | 80-89 | Pre-hypertension |
| | 90+ | High Hypertension |
| **Smoking rate (Cig/Day) [25]** | 1 - 10 | Light |
| | 11 - 19 | Moderate |
| | 20+ | Heavy |

*1) Calculating the OP's Priority using MILP-Compliant Naïve Bayesian Formulation*

We developed the following formulations to include the naïve Bayesian classifier within the MILP model, where it calculates the likelihood $PS_z$ of a stroke given a certain *current state* $CS_i$. The model then transforms this likelihood into an updated *user priority (weight)* $UP_k$ indicated in equation (7).

Rewriting equation (1) in a mathematical programming formulation gives:

$$CP_{i,v}^{c,z} = P\left(F_i = V_{F_i}^{j,z} \middle| C_i = V_{C_i}^{r,z}\right) = \sum_{d=1}^{|D|} \sum_F \sum_C \frac{S_{F_i C_i}^{j,r,d,z}}{G_{C_i}^{r,d,z}} \quad (3)$$

$$\forall i \in \mathcal{I}, c \in \mathcal{C}, z \in \mathcal{Z}$$

Where equation (3) is used to calculate the conditional probability $P(F_i|C_i)$ in the MILP model. The nominator represents the total number of days where the outpatient $z$ has a certain value $V_{F_i}^{j,z}$ that we want to test, *and* a stroke (indicated by $V_{C_i}^{1,z}$) where $C_1$ depicts the class stroke and $r = 1$ registers the stroke occurrence. The denominator represents the total number of stroke days.

$$S_{F_i C_i}^{j,r,d,z} \geq 0 \quad (4)$$

$$\forall z \in \mathcal{Z}, i \in \mathcal{I}, d \in \mathcal{D}$$

$$S_{F_i C_i}^{j,r,d,z} = E_{F_i}^{j,d,z} + G_{C_i}^{r,d,z} - 1 \quad (5)$$

$$\forall z \in \mathcal{Z}, i \in \mathcal{I}, d \in \mathcal{D}$$

Equations (4) and (5) achieve a logical AND operation in which the binary variable $S_{F_i C_i}^{j,r,d,z} = 1$ when both binary variables $E_{F_i}^{j,d,z}$ and $G_{C_i}^{r,d,z}$ are equal to 1. This variable indicates that outpatient $z$ with the $j^{th}$ value of feature $F_i$ has the $r^{th}$ value of class $C_i$ in day $d$.

Rewriting equation (2) gives:

$$PS^{z,r} = \left[\sum_{d=1}^{|D|} \frac{G_{C_i}^{r,d,z}}{|D|}\right] \prod_{i=1}^{\mathcal{I}} P\left(F_i = V_{F_i}^{CS_i,z} \middle| C_i = V_{C_i}^{CS_i,z}\right) \quad (6)$$

$$\forall z \in \mathcal{Z}$$

Equation (6) represents the formulation we used to determine the probability of stroke $PS^{z,r}$. Given a *current state* $CS_i$, all feature variables $F_i$ are considered. This means $i$ has the range $i \leq |\mathcal{I}|$ (in this work $i = 1, .., 4$). The L.H.S. represents the posterior probability that outpatient $z$ has a stroke. The first term on the R.H.S. represents the prior probability of stroke and the second term on the R.H.S. represents the joint probability that patient $z$ has the given values of the features. The multiplication of the two terms on the R.H.S. shows the naïve nature of the Naïve Bayesian estimate in this case where the features are assumed independent.

$$UP_k = 1 + \alpha \cdot PS^{z,r} \quad (7)$$

$$\forall k \in \mathcal{K} : z = k, k \succ NU$$

The user weight $UP_k$ is calculated as shown in equation (7). Since the naïve Bayesian classifier produces probabilities of small magnitude, we multiplied the overall probability of stroke ($PS^{z,r}$) by a tuning factor $\alpha$ to produce an effective-yet-reasonable weight, which drives the objective function into favoring the imperiled outpatients.

*C. Problem Formulation*

Using our track record in MILP optimization and heuristics formulation in [26-34], and physical layer modelling track record in [35-39], we developed the following MILP models to optimize the cellular system resource allocation for OPs and normal users. We consider the OPs monitoring system to operate in a scenario of an LTE-A network comprising $B$ base stations represented by set $\mathcal{B} = \{1, ..., B\}$, operating at channels with 1.4 MHz bandwidth. Each base station $b$ has $N$ PRBs represented by set $\mathcal{N} = \{1, ..., N\}$. The network serves $K$ users (normal and OPs) represented by set $\mathcal{K} = \{1, ..., K\}$ by allocating PRB $n$ to connect to BS $b$ in an instant in time. The goal is to optimize the uplink of the LTE-A network, so that the OPs are prioritized over normal users; thus, allocating them high-powered PRBs.

We formalize this problem as a MILP model. Table III defines the sets, parameters, and variables used in the network optimization problem formulation.

TABLE III

SYSTEM SETS, PARAMETERS, AND VARIABLES

| **Sets** | |
|---|---|
| $\mathcal{K}$ | Set of users. |
| $\mathcal{N}$ | Set of physical resource blocks. |
| $\mathcal{B}$ | Set of base stations. |
| $\mathcal{D}$ | Set of days. |
| $\mathcal{F}$ | Set of features in learning dataset. |
| $\mathcal{C}$ | Set of classes in learning dataset. |
| $V_{F_i}^r$ | Set of values feature $F_i$ can take in the learning dataset. |
| $V_{C_i}^r$ | Set of values a class variable $C_i$ can take in the learning dataset. |
| $\mathcal{I}$ | Set of features and class variables. |
| $\mathcal{Z}$ | Set of outpatient users, $(\mathcal{Z} \subset \mathcal{K})$. |

| **Parameters** | |
|---|---|
| $CP_{i,v}^{c,z}$ | Conditional probability that input feature $i$ takes the value $v$ given that outpatient $z$ has class $C$ considering input feature $i$ of value $v$ given class $c$ for outpatient $z$. |
| $CS_i$ | Current state of the patient in feature $i$ (e.g. Cholesterol value). |
| $V_{F_i}^{CS_i,z}$ | $CS_i{}^{th}$ value taken by feature $F_i$ for patient $z$. |
| $V_{C_i}^{CS_i,z}$ | $CS_i{}^{th}$ value taken by class $C_i$ for patient $z$. |
| $E_{F_i}^{j,d,z}$ | Binary variable, $E_{F_i}^{j,d,z} = 1$ if feature $F_i$ takes the $j^{th}$ value on day $d$ for outpatient $z$, 0 otherwise. |
| $G_{C_i}^{r,d,z}$ | Binary variable, $G_{C_i}^{r,d,z} = 1$ if class $C_i$ takes the $r^{th}$ value on day $d$ for outpatient $z$, 0 otherwise. |
| $S_{F_i C_i}^{j,r,d,z}$ | Binary variable, $S_{F_i C_i}^{j,r,d,z} = 1$ if $E_{F_i}^{j,d} = 1$ and $G_{C_i}^{r,d} = 1$ (Logical AND operation). |
| $UP_k$ | User priority ($UP_k$ =1 for normal users whereas $UP_k > 1$ is granted for OPs depending on their risk factor). |
| $Q_{k,n}^b$ | Power received from user $k$ using physical resource block $n$ at base station $b$. |
| $H_{k,n}^b$ | Rayleigh fading with zero mean and a standard deviation equal to 1 experienced by user $k$ using physical resource block $n$ at base station $b$. |
| $A_k^b$ | Signal attenuation experienced by user $k$ using physical resource block $n$ at base station. |
| $PM$ | Maximum power allowed per uplink connection. |
| $P$ | Power consumed to utilize physical resource block $n$ to connect user $k$ to base station $b$. |
| $\lambda$ | An arbitrary, large positive value. |
| $\sigma_{k,n}^b$ | Additive White Gaussian Noise (AWGN) power in watts experienced by user $k$ using physical resource block $n$ at base station $b$. |
| $PS^{z,r}$ | Probability of stroke of outpatient $z$. |
| $m_{y,k}$ $h_{y,k}$ | Piecewise linearization equation coefficients for line $y$ of user $k$. |
| $\alpha$ | Tuning factor. |
| $NU$ | The total number of normal users. |
| **Variables** | |
| $X_{k,n}^b$ | Binary decision variable $X_{k,n}^b = 1$ if user $k$ is assigned physical resource block $n$ in base station $b$, otherwise $X_{k,n}^b = 0$. |
| $T_{k,n}^b$ | The SINR of user $k$ utilizing physical resource block $n$ at base station $b$. |
| $\phi_{m,n,k}^{w,b}$ | Non-negative linearization variable where $\phi_{m,n,k}^{w,b} = T_{k,n}^b X_{m,n}^w$. |
| $S_k$ | SINR of user $k$. |
| $L_k$ | Logarithmic SINR of user $k$. |

A user's SINR at the uplink side of an OFDMA network can be expressed as [40].

$$T_{k,n}^b = \frac{Signal}{Interference + Noise} = \frac{Q_{k,n}^b X_{k,n}^b}{Q_{m,n}^b X_{m,n}^w + \sigma_{k,n}^b} \quad (8)$$

Examining the numerator (i.e. signal), $Q_{k,n}^b X_{k,n}^b$ represents the signal power received at the BS side from user $k$. The binary decision variable $X_{k,n}^b = 1$ indicates that user $k$ is connected to BS $b$ and occupies PRB $n$. The power received at BS $b$ from the interfering user(s) $m, m \neq k$, on the same PRB is $Q_{m,n}^b X_{m,n}^w$; while $X_{m,n}^w$ indicates that the interfering user(s) $m$ is connected to another BS $w, w \neq b$ on PRB $n$. The Additive White Gaussian Noise (AWGN) is annotated as $\sigma_{k,n}^b$. A graphical illustration of equation (8) is shown in Fig. 3.

Rewriting equation (8):

$$\sum_{\substack{w \in \mathcal{B} \\ w \neq b}} \sum_{\substack{m \in \mathcal{K} \\ m \neq k}} T_{k,n}^b Q_{m,n}^b X_{m,n}^w + T_{k,n}^b \sigma_{k,n}^b = Q_{k,n}^b X_{k,n}^b$$
$$\forall \, k \in \mathcal{K}, n \in \mathcal{N}, b \in \mathcal{B} \quad (9)$$

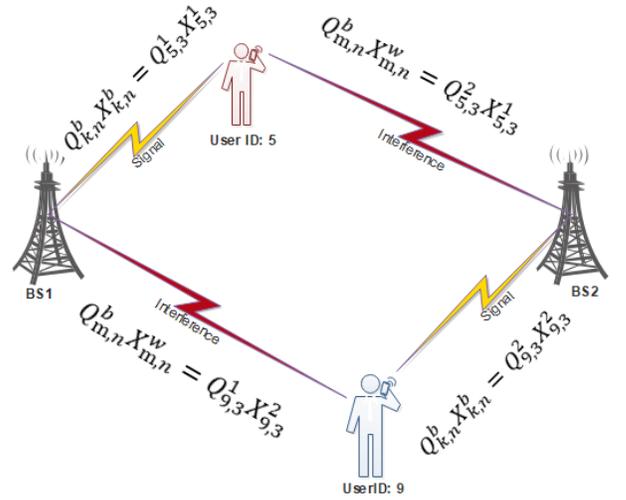

Fig. 3. User Interference

The first term in (9) is nonlinear (quadratic) as it involves the multiplication of two variables (Continuous $T_{k,n}^b$ and Binary $X_{m,n}^w$). Therefore, linearization is essential to solve the NP-hard model using a linear solver such as CPLEX, where the linearization is given in (12) to (15).

We have developed two approaches to solve the resource allocation problem. The first approach uses an objective function that maximizes the Weighted Sum-Rate of the SINRs experienced by the users. The second approach introduces fairness among the users by employing a Proportionally Fair (PF) objective function.

*1) MILP Formulation for the WSRMax approach*

The objective is to maximize the system's overall SINR. This can be realized through the maximization of the individual users' SINRs.

### a) Before Prioritizing the OPs

The OPs' risk factors introduced in the previous section are scaled into priorities (i.e. weights) and used to prioritize the OPs over other users. The MILP model is formulated as follows:

**Objective:** Maximize

$$\sum_{k \in \mathcal{K}} \sum_{n \in \mathcal{N}} \sum_{b \in \mathcal{B}} T_{k,n}^b \, UP_k \tag{10}$$

The objective given in (10) aims to maximize the weighted sum of the users' SINRs. These weights (i.e. priorities) are higher for OPs compared to healthy users and proportional to the OPs calculated risk factor. Note that $UW_k$ have an initial value of 1 for all users as shown in (11). However, the OPs will have updated values according to their risk factor. This will ultimately drive the system into prioritizing the OPs over the healthy users during PRB assignment. The mathematical formulations related to the OP weight (priority) calculation was illustrated in subsection B.1.

All users have equal priorities (i.e. weights) at this stage as shown in (11).

$$UP_k = 1 \tag{11}$$
$$\forall \, k \in \mathcal{K}$$

**Constraints:**

Equations (12) - (15) represent the constraints required to multiply a float by a binary variable while maintaining linearity [40]:

*Subject to:*

$$\phi_{m,n,k}^{w,b} \geq 0 \tag{12}$$

Replacing the quadratic term $T_{k,n}^b X_{m,n}^w$ with the linearization variable $\phi_{m,n,k}^{w,b}$ that incorporates all the indexes of the multiplied variables.

$$\phi_{m,n,k}^{w,b} \leq \lambda X_{m,n}^w \tag{13}$$
$$\forall \, k,m \in \mathcal{K}, n \in \mathcal{N}, w, b \in \mathcal{B}, (m \neq k, b \neq w)$$

$$\phi_{m,n,k}^{w,b} \leq T_{k,n}^b \tag{14}$$
$$\forall \, k,m \in \mathcal{K}, n \in \mathcal{N}, w, b \in \mathcal{B}, (m \neq k, b \neq w)$$

$$\phi_{m,n,k}^{w,b} \geq \lambda X_{m,n}^w + T_{k,n}^b - \lambda \tag{15}$$
$$\forall \, k,m \in \mathcal{K}, n \in \mathcal{N}, w, b \in \mathcal{B}, (m \neq k, b \neq w)$$

After replacing $T_{k,n}^b X_{m,n}^w$ with $\phi_{m,n,k}^{w,b}$, equation (9) can thus be rewritten as in (16). $\phi_{m,n,k}^{w,b} = T_{k,n}^b X_{m,n}^w$ is equal to the SINR of user $k$ connected to base station $b$ with physical resource block $n$ if there is an interfering user $m$ connected to the other base station $w$ with the same physical resource block $n$; it is zero otherwise.

$$\sum_{\substack{w \in \mathcal{B} \\ w \neq b}} \sum_{\substack{m \in \mathcal{K} \\ m \neq k}} Q_{m,n}^b \phi_{m,n,k}^{w,b} + T_{k,n}^b \sigma_{k,n}^b = Q_{k,n}^b X_{k,n}^b \tag{16}$$
$$\forall \, k \in \mathcal{K}, n \in \mathcal{N}, b \in \mathcal{B}$$

$$\sum_{n \in \mathcal{N}} P \, X_{k,n}^b \leq PM \tag{17}$$
$$\forall \, k \in \mathcal{K}, b \in \mathcal{B}$$

Constraint (17) ensures that the users do not exceed their maximum available amount of power per uplink connections (in case more than one PRB is utilized by the same user $k$). In the current work, the user is allowed a single PRB.

$$\sum_{k \in \mathcal{K}} X_{k,n}^b \leq 1 \tag{18}$$
$$\forall \, n \in \mathcal{N}, b \in \mathcal{B}$$

Constraint (18) limits the assignment of each PRB to one user only.

$$\sum_{b \in \mathcal{B}} \sum_{n \in \mathcal{N}} X_{k,n}^b \geq 1 \tag{19}$$
$$\forall \, k \in \mathcal{K}$$

Constraint (19) guarantees that each user is assigned at least one PRB from any BS. Thus, no user is left without service. Additionally, this prevents the MILP from blocking interfering users to maximize the total SINR.

### b) After Prioritizing the OPs

In this approach, OPs' risk factors introduced in the previous section are scaled into weights to prioritize the OPs over other users. The MILP model is formulated in the same way as mentioned in the previous subsection. However, equation (7) is included in this model to represent the OPs' weights (i.e. priorities) while (11) is replaced by (20) to cover the normal users only.

$$UP_k = 1 \tag{20}$$
$$\forall \, k \in \mathcal{K}: 1 \leq k \leq NU$$

## 2) MILP Formulation for the PF Approach

In this approach, the objective is to maximize the logarithmic sum of the user's SINRs. Due to the nature of the natural logarithm, a slight decrease in the overall SINR might be observed but to the expense of preserving fairness among normal users.

### a) Before Prioritizing the OPs

In this case, all users are treated equally, thus there is no prioritization in terms of resource allocation. However, keeping fairness among users still holds as a necessity. Since the only part that we are dealing with is the value of the individual user's SINR, and to simplify the manipulation of the equation before adding the natural logarithm part, we present the optimization variable $S_k$, to serve as the SINR for each user $k$.

$$S_k = \sum_{n \in \mathcal{N}} \sum_{b \in \mathcal{B}} T_{k,n}^b \tag{21}$$
$$\forall \, k \in \mathcal{K}$$

Equation (21) replaces the three-indexed variable $T_{k,n}^b$ with a single-indexed variable $S_k$.

$$L_k = \ln S_k \tag{22}$$
$$\forall \, k \in \mathcal{K}$$

Equation (22) calculates $L_k$ as a logarithmic function of the user's SINR $S_k$. Since the natural log is a concave function, and to preserve the linearity of our model, piecewise linearization was used as depicted in constraint (24).

The objective is as shown in (23):

**Objective**: Maximize

$$\sum_{k \in K} L_k \tag{23}$$

**Constraints:**

In addition to constraints (12)-(19) from the previous model, the PF satisfies the following constraint

*Subject to:*

$$L_k \leq m_{y,k} * S_k + h_{y,k} \tag{24}$$
$$\forall \, k \in \mathcal{K}$$

Constraint (24) represents a set of piecewise linearization relations implemented to linearize the concave function in equation (22). Note that constraint (24) corresponds to the line equation $y = mx + h$ where $m$ and $h$ denote the equation coefficients.

### b) After Prioritizing the OPs

In this case, the outpatients are prioritized. Equation (22) is rewritten to reflect the change.

$$L_k = \ln S_k \tag{25}$$
$$\forall \, k \in \mathcal{K} : 1 \leq k \leq NU$$

Equation (25) shows that the log function is applied to the normal users only. The OPs, on the other hand, are assigned weights instead.

**Objective**: Maximize

$$\sum_{k \in K, 1 \leq k \leq NU} L_k + \sum_{k \in K, k > NU} S_k U P_k \tag{26}$$

The multi-objective function in (26) (i) maximizes the sum of the SINRs allocated to all users, (ii) Assigns OPs priority by allocating OPs PRBs with high SINRs that reflect their relative priority, and (iii) Implements Fairness: by assigning healthy users PRBs with comparable SINRs. These objectives were implemented by adding both the summation of a log function of the healthy users' SINRs (i.e. Proportional Fairness) and the weighted sum of the OPs' SINRs (OPs priority).

**Constraints:**

The model satisfies constraint (12) - (19) from the previous approach. In addition to equation (20) and:

$$L_k \leq m_{y,k} * S_k + h_{y,k} \tag{27}$$
$$\forall \, k \in \mathcal{K}, k \leq NU$$

Constraint (27) represents the same set of equations for the piecewise linearization that was used in constraint (24), however, the difference is in the range of users it is applied to.

### 3) Calculating the Received Power

The received signal power (in Watts) $Q_{k,n}^b$ varies according to the channel conditions and the distance between the user and the BS. Considering Rayleigh fading denoted by $H_{k,n}^b$ and distance dependent path loss denoted by $A_{k,n}^b$, the received signal power is given as:

$$Q_{k,n}^b = P \, H_{k,n}^b A_k^b \tag{28}$$

where $H_{k,n}^b$ denotes Rayleigh fading and $A_k^b$ represents power loss due to attenuation (distance dependent path loss) and is given by [41]:

$$A \, (dBm) = 128 + 37.6 \, \log_{10} \frac{distance(meters)}{1000} \tag{29}$$

To unify the units, equation (30) is used to convert the power to Watts.

$$A \, (mw) = 10^{\frac{A(dBm)}{10}} \tag{30}$$

## IV. Heuristic

To provide a method to validate the MILP operation and to deliver a real time solution, a heuristic approach was developed to optimize the PRBs assignment based on the user's priority. The heuristic, as shown in the flowchart in Fig. 4, starts by initializing the data parameters, sets, variables and reads the received power (Q) values from a separate file. A check for user prioritization takes place. This affects the users' admittance order to the system. If user prioritization is ON (i.e. big data analytics are used), the OPs will be arranged according to their priority such that the most critical OP will be served first. This kind of check is vital at this stage due to the sequential nature of the heuristic, thus, the first few users will be granted high SINRs due to the higher number of available channels. OPs do not compete with each other over the available PRBs, i.e. their interfering candidates are normal users only. Finding the PRB at which a user achieves a relatively-high SINR is done by assigning a PRB where interference is attributed to a subset of $|\mathcal{B}|$-1 interferers with minimum interfering power to that user at its PRB, where $|\mathcal{B}|$ is the number of BSs (the cardinality of $\mathcal{B}$). As the heuristic continues to run, the PRB availability is reduced. Once the PRBs are allocated to the OPs, the total number of allocated PRBs will equal to $(2 * \mathcal{Z})$. On the other hand, the number of free PRBs (FPRB) will be equal to $\lceil \mathcal{B} * \mathcal{N} \rceil - [2 * \mathcal{Z}]$ giving a total of $2^{FPRB}$ combinations. Finding an interfering user with the minimum power on each RB (i.e. maximum SINR) results in reducing the above number of combinations. Accordingly, a pool with the length $|FPRB|$ comprised of the highest achievable SINR on each PRB will be formed. The heuristic follows a semi-greedy approach [42]. Thus, one SINR will be randomly selected from the pool of best SINRs. The reasons behind this selection criterion are (i) to establish local fairness between the user and its interferer so that the interferer does not endure a huge impact by being assigned a very low-powered PRB; moreover, (ii) to conform to the objective function in which each individual user's SINR is maximized while maximizing the overall system-wide SINR. Once the user is assigned a SINR, the corresponding PRB(s) is assigned to the user and the interferer. The heuristic repeats the above procedure for the remaining users. Due to its sequential nature, this heuristic was iterated 1000 times, randomizing the users' admission order (serving sequence) to the system in each iteration, while maintaining the semi-deterministic nature of the interferer's PRB assignment stage. The users' average SINRs are then calculated. Thus, applying this heuristic over different realizations of the network instates fairness among users in the long run. Sensitivity analysis was carried out to calculate the 95% confidence interval. To that end, the heuristic was applied to over 100 files each containing different values representing the powers received from the BS. Concurring results between the heuristic and the MILP model operation can be observed, as will be shown in the results section.

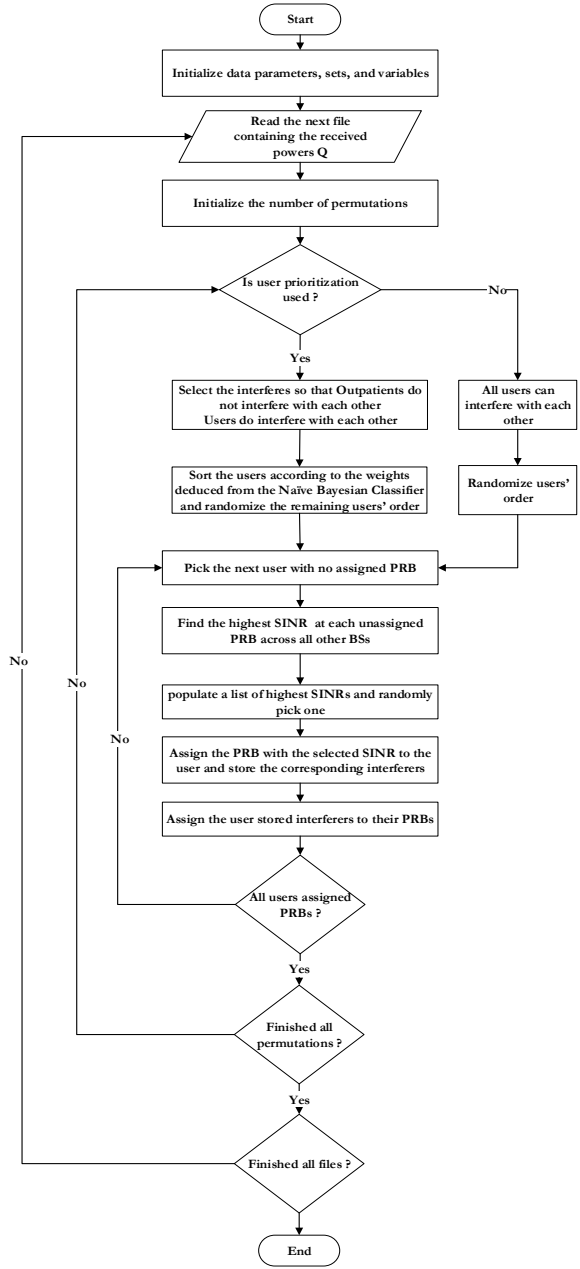

Fig. 4. The heuristic flowchart

## V. RESULTS AND DISCUSSION

Before delving into the results of the MILP model and heuristic, the parameters indicated in Table IV should be noted. We consider a cellular network that operates in an urban environment, hence Rayleigh fading channel model with path loss. The results evaluate two scenarios; the first represents the state of the network before using big data analytics to prioritize the OPs.

In this case, all the users were given equal base priority (i.e. weight) of 1. The second scenario represents the network state after using big data analytics where the OPs' priorities are updated according to their risk factor and the value of the tuning factor $\alpha$.

The OPs' stroke likelihood $PS^{z,r}$ were 0.0032, 0.0064, and 0.00208 for users 8, 9, and 10, respectively. The use of equation (7) produced $1.104 \leq UP_k \leq 1.32, 1.208 \leq UP_k \leq 1.64,\ 1.312 \leq UP_k \leq 1.96m, 1.52 \leq UP_k \leq 2.6, 2.04 \leq UP_k \leq 4.2$ user priorities according to tuning factor values of $\alpha$ of 50, 100, 150, 250 and 500, respectively.

TABLE IV

MODEL PARAMETERS

| Parameter | Description |
|---|---|
| LTE-A system bandwidth | 1.4 MHz |
| Channel Model | Path Loss [41] and Rayleigh fading [40] |
| No. of BS | 2 |
| Number of PRBs per BS | 5 |
| Number of users | 10 |
| Number of normal users ($NU$) | 7 |
| Number of OPs | 3 |
| AWGN ($\sigma_{k,n}^b$) | -162 dBm/Hz [41] |
| Distance between user $k$ and BS $b$ | (300 - 600) m |
| Maximum transmission power per connection $PM$ | 23 dBm [41] |
| UE transmission power per PRB | 17 dBm |
| Base (i.e. normal user priority) weight | 1 |
| Outpatient priority $UP_k$ calculation method | Naïve Bayesian classifier |
| OP observation period | 30 Days |
| $\alpha$ values | 50, 100, 150, 250, and 500 |

### A. The WSRMax Approach

#### 1) Before Prioritizing the OPs

In this scenario, big data analytics is not employed to prioritize the OPs, i.e., all users have equal weights equivalent to the *base user weight* (i.e. 1). Observing Fig. 5, it can be seen that the OPs (represented by users 8, 9, and 10, in both the MILP and heuristic results) are assigned PRBs with near average SINR as the MILP and heuristic strive to maximize the overall SINR.

Analogous SINR values can be observed in Fig.5 for both the MILP and the heuristic. The average SINRs computed through the heuristic and the MILP approaches are comparable at around 5.4 and 5.5, respectively.

As a measure of fairness, i.e. to quantify how close the SINR values are to the mean, we considered accentuating the Standard Deviation (SD) for the users' SINRs. The results are 0.4 and 0.3 for the heuristic and the MILP, respectively. Thus, the results confirm that the heuristic can approach the MILP and provide an acceptable level of fairness among the users by implementing the described permutation over independent realizations of the channel, at the expense of slightly sacrificing the overall SINR. An extensive sensitivity analysis was carried out, and 95% confidence intervals for each user's SINRs are depicted in Fig. 5. The average

SINR lied between 5.1 and 6 for the MILP results, and between 4.5 and 5.7 for the heuristic results.

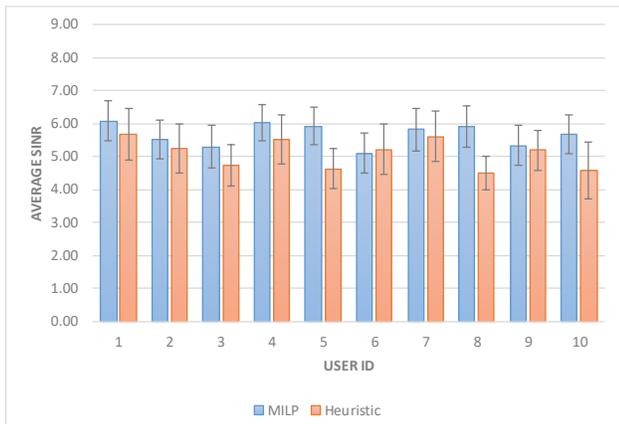

Fig. 5. Users' SINR before using big data analytics (WSRMax Approach)

### 2) After Prioritizing the OPs

In this scenario, the use of big data analytics resulted in assigning OPs higher priority than normal users by means of the naïve Bayesian classifier. The results shown in Fig. 6 clearly demonstrate that all the OPs (users 8, 9, and 10) were assigned PRBs with high SINRs compared to their previous SINRs in Fig. 5. The system wide performance is a trade-off (*optimally* selected) between the task of assigning higher SINRS to OPs versus a reduction in the average SINR in this scenario (between 0.3% ($\alpha$=50) and 6% ($\alpha$=500)) compared to the average SINR in the first scenario. This reduction in the average SINR is due to the fact that the system was forced to choose a PRB assignment scheme that prioritizes the maximization of OPs' individual SINRs over the total SINR. The results also show that the heuristic approaches the MILP performance, with a very comparable SINRs, however, the heuristic mostly displayed a marginally higher OP SINRs. This is due to the sequential nature of the heuristic which forced the system to serve the OPs first after further arranging them according to their priorities. This challenge was mitigated by preparing a list of highest achievable SINRs and randomly selecting one. The selection criterion of the user and its interferer was conducted on a sequential and semi-deterministic manner, respectively to instate fairness between users as illustrated in section IV.

The results in Fig. 6 depict an agreement in terms of the average SINR between the heuristic (5.1) and the MILP (ranged from 5.3 to 5.6 depending on the value of $\alpha$). This approach slightly impacted the fairness between normal users as will be shown in the upcoming subsection.

The average SINR of an individual user ranged between 4 and 7.6 for the MILP ($\alpha$=500), and between 3.7 and 7.9 for the heuristic. A clearer illustration can be observed in Fig. 6 where the confidence interval for each individual user's SINRs is shown.

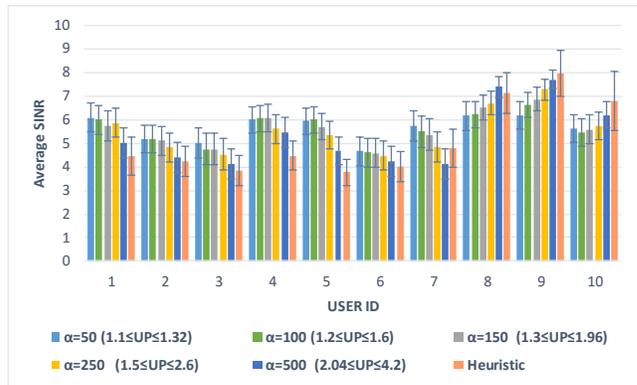

Fig. 6. Users' SINR after user prioritization (WSRMax Approach)

### 3) The impact of $\alpha$ on Fairness and SINR

The proposed model can be fine-tuned using the parameter $\alpha$ (i.e. tuning factor) introduced in equation (7). This parameter enables the reciprocity between the achievable fairness among users quantified by the SD and the average SINR. We examined the effect on the average SINR and the SD of using different values of $\alpha$ as illustrated in Fig. 7 and in Fig. 8. Increasing the value of $\alpha$ directs the system to focus more on the OPs; consequently, a trade-off takes place resulting in lower values of the system's average SINR as seen in Fig 8, to increase the SINR of the selected users (i.e. the OPs), negatively affecting fairness as illustrated by the increasing SD in Fig. 7.

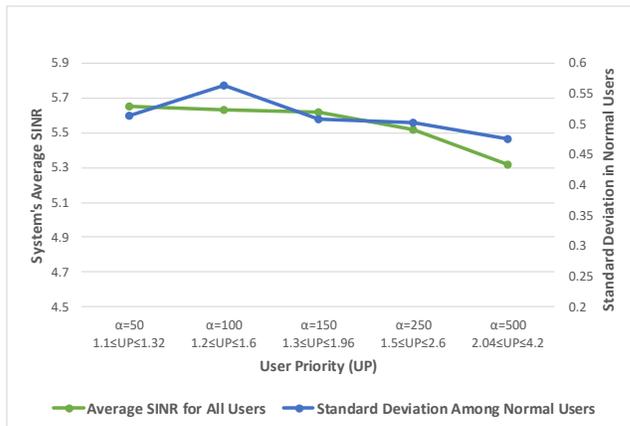

Fig. 7. The effects of changing $\alpha$ on fairness and average SINR (WSRMax Approach)

It should be noted that the individual SINRs for the OPs correspond to the weights given to each OP using the Naïve Bayesian Classifier. Sorting the users according to these weights produces an order that conforms to the values depicted in Fig. 8. The highest SINR was granted to user 9 which is the OP with highest probability of stroke; thus the highest priority, while the lowest among the three OPs was user 10 who also happened to be the one with the least priority among the OPs (nevertheless still higher than the normal users).

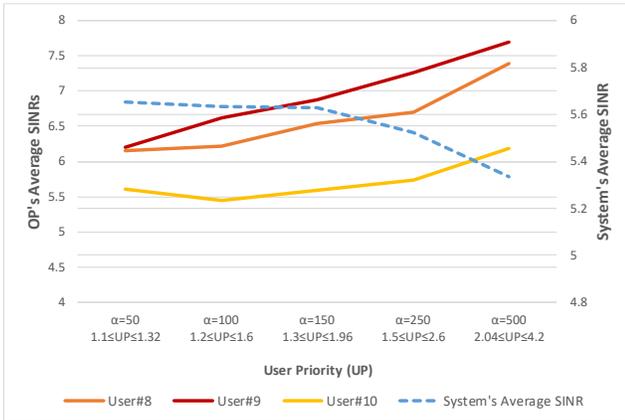

Fig. 8. The impact of α on both user and average SINRs (WSRMax)

### B. The PF Approach

#### 1) Before Prioritizing OPs

The objective function in (23) is applied to this scenario. The goal is to maximize the summation of the log of the users' SINRs while ensuring fairness without prioritizing a certain subset of users. The results shown in Fig. 9 bares a trend similar to the one depicted in Fig. 5. However, due to the nature of the log function used in the objective function, fairness was maintained between the users (SD of 0.3 and 0.4 for the MILP and the heuristic, respectively), while the total SINR was reduced by 7% compared to the one produced by the MILP in the WSRMax approach. The average SINRs for the heuristic and the MILP approaches are analogous at around 5.1 and 5.3, respectively. Sensitivity analysis was performed (95% confidence interval) where the average SINR achieved by the MILP ranged between 4.4 and 6.1, and between 4.1 and 6.4 for the heuristic results.

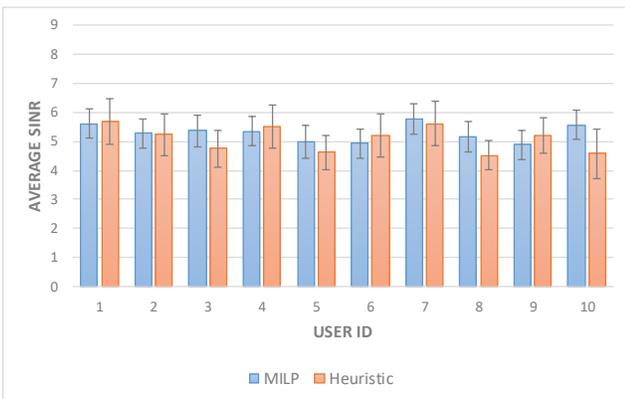

Fig. 9. Users' SINR before user prioritization (PF Approach)

#### 2) After Prioritizing OPs

In this scenario, the OPs' priorities (i.e. weights) are updated according to the stroke likelihood determined through the use of big data analytics. The objective function in (26) is used; consequently, the model grants the OPs high powered PRBs as can be noted in Fig. 10. Comparing the PF approach to the WSRMax approach, it is evident that this approach grants the OPs higher SINRs (traded off with the other users). Furthermore, this approach shows higher conformance between the heuristic and MILP than the previous one. However, this was accomplished by trading off the average SINR. The MILP scored an average SINR between 5.2 ($\alpha = 50$) and 4.9 ($\alpha = 500$) as can be seen in Fig. 10, while the heuristic's average SINR is 5.1.

Narrower confidence intervals can be noted when employing this approach. As a matter of fact, this is a good indication of the precision of the approach in hand, thus producing results with narrower margins of error than the previous approach.

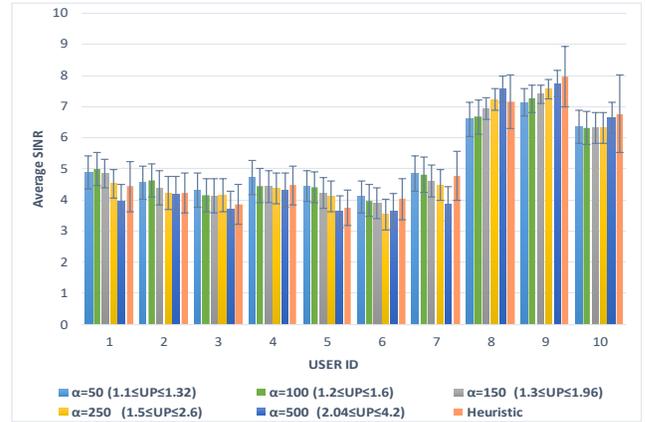

Fig. 10. Users' SINR after user prioritization (PF Approach)

#### 3) The Impact of α on Fairness and SINR

Increasing the weights allocated to the OPs in this approach has similar effects to the ones in the previous subsection V.A.3 as shown in Fig. 11 and in Fig. 12. The reduction in the SINR is around 4%. However, the OPs were assigned higher SINRs. Furthermore, better fairness was reported among healthy users with a SD between 0.27-0.32 (depending on the value of $\alpha$). Thus, offering a more stable approach.

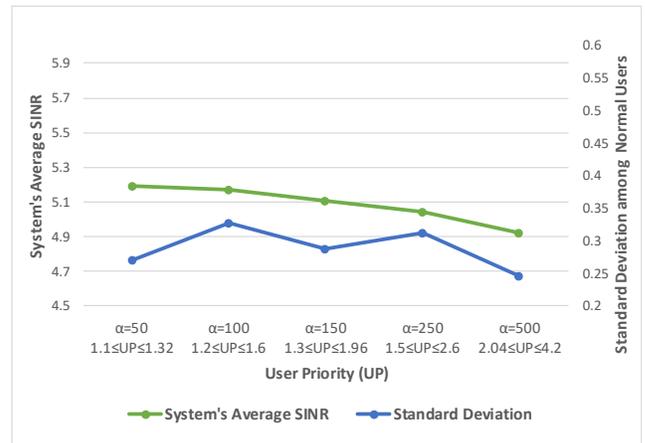

Fig. 11. The effects of changing α on fairness and average SINR (PF Approach)

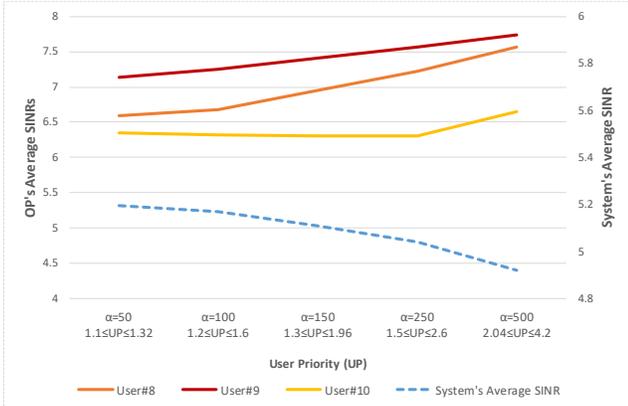

Fig. 12. The impact of α on both user and average SINRs (PF Approach)

Further analysis of Fig. 6 and Fig. 10 reveals that the SINR sum achieved by the WSRMax approach is larger than that of the PF approach. Since the WSRMax target is to maximize the sum rate (which is what an unregulated operator tries to do) while the PF approach introduces fairness, hence resources are not all allocated to the user with the best channel. The PF approach improves fairness, but reduces the sum rate (which is the case of a regulated operator).

### C. *Testing the Heuristic's Scalability*

Employing higher LTE-A system bandwidths enables the operator to serve more users creating a challenge for the developed heuristic to allocate resources to OPs with minimum delay to serve their urgent needs. To evaluate the scalability of the heuristic the heuristic elapsed time is considered.

We considered a scenario with six cases where the system operates at bandwidths of 1.4, 3, 5, 10, 15, and 20 MHz and increased the number of users, where all PRBs are occupied. For each case we measured the time it takes the heuristic to allocate all users appropriate PRBs. The heuristic elapsed time was measured using the MATLAB functions *tic* and *toc*. Time calculation was carried out on a Windows 10 computer equipped with Intel core i5-4460 3.2 GHz quad core processor and 16 GB of RAM. Fig. 13 shows the heuristic's total elapsed time (in seconds) versus the number of users on the $x$ axis.

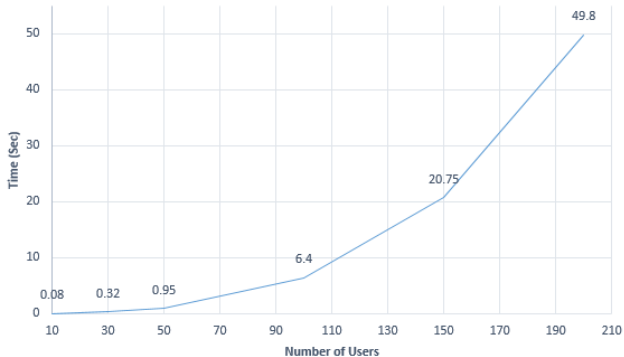

Fig. 13. The Heuristic's Scalability

The proposed heuristic tries to serve $K$ users to be allocated to $K/2$ PRBs on each of the two BSs with another loop dedicated to interferer allocation. The first run contains a search of total $K$ possible interferers (before satisfying the condition $k \neq m$). This means it requires $O(N * \frac{N}{2} * 2 * N)$ time. Additionally, the MATLAB sort function requires $O(N \log N)$ time [43]. Thus, the overall complexity is $O(N^4 \log N)$. The proposed heuristic provided a reduction in the run time to solve the NP-Hard problem [40] with a slight sacrifice in the accuracy of the results.

## VI. Open Research Directions

### A. *Choosing the Decision-making Entity*

Choosing the optimal type and location of computing (e.g. cloud, fog, etc.) is a separate optimization problem. Additionally, this may depend on other factors (or variables) like the ratio of OPs to normal users.

### B. *Testing the impact of the Feature Ranking Techniques*

The current system treats the feature variables on an equal basis. However, we plan to further-study the impact of each feature and correspondingly employ a suitable feature ranking technique. The impact of this technique can then be verified with clinical help.

### C. *Routing within Small Cells in 5G Networks with Privacy*

The proposed solution can be integrated in 5G networks. Optimized routing algorithms can be developed to carry the OPs' traffic through the small cells with minimum latency. In addition, it is vital to protect the OPs' privacy through the traversed hops. This can be addressed by classifying the OPs' data in a ranking system, where the highest rank is treated as the most private medical data. Hence, a specific (secure) route is selected.

### D. *Impact of OP Mobility*

Grouping the OPs into clusters with common mobility patterns allows the operator to know in advance if there are some areas with high OP density. Hence, prepare the network. This means deploying more nodes so that these OPs do not severely impact the network operation. In addition, our current system works on a given realization of the patient data and channel conditions (although consideration is given to many realizations). However, in a real-world scenario, there is a constant change in the number of users accessing and leaving the BS coverage. Such dynamic behavior should be addressed, possibly by OP weighted beamforming and Beamsteering.

### E. *Use of Infrastructure Sharing and Game Theory*

The use of infrastructure sharing can help ensure the widest coverage since the resulting area is the combination of all the local (or national) operators' coverage at reduced cost. To encourage the operators to participate, game theory can be used to establish coalitions, such that, for example, the higher the number of OPs, the more revenue is awarded to the operator, e.g., reduced taxes.

### F. *Wireless energy transfer for Remote Drug Injection*

Ensuring high-energy transfer in the downlink might be integrated with our approach to power the body sensors or to actuate a drug-injection mechanism. This can be used in the case of a sudden degradation in the health parameters especially in the case of critical conditions such as diabetes. The reliability of such an approach should be evaluated and improved.

## VII. Conclusions

This paper introduced a system that employs the power of big data analytics to optimize the uplink of an LTE-A cellular network. OP's medical record and readings from medical IoT sensors are processed in a big data analytics engine to find the likelihood of a stroke for an OP. The goal is to target OP users within the network to ensure they can always have access to the best wireless resources when in need. The proposed system achieves that with minimal impact on the wireless system-wide performance and SINR levels among healthy users in the network, thus improving the network utility for telecom operators while saving human lives and preserving fairness among normal users. Two approaches (WSRMax and PF) were presented and compared in terms of the average SINRs and fairness. The WSRMax approach improved the OPs' average SINR by up to 26.6%, whereas for PF approach increased them by 40.5%. The average SINR for normal users ranged between 5.5 and 4.6 using the WSRMax approach while the PF approach reported a range between 4.6 and 4 (depending on $\alpha$). Fairness among users was quantified using SD. The WSRMax approach granted the healthy users SINRs with a SD between 0.47 and 0.56 (depending on $\alpha$) while the PF approach ranged between 0.24 and 0.3 SD. Furthermore, we developed a real-time heuristic to verify the MILP operation. The heuristic achieved comparable results to the MILP, and we demonstrated the heuristic's scalability. We also presented several open research directions that we believe, if appropriately addressed, would ultimately refine the way future cellular networks can react to their users' needs.

## VIII. Acknowledgement

The authors would like to acknowledge funding from the Engineering and Physical Sciences Research Council (EPSRC), INTERNET (EP/H040536/1) and STAR (EP/K016873/1) projects. All data are provided in full in the results section of this paper.

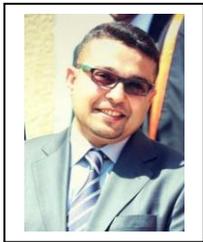

**Mohammed S. Hadi** received the B.Sc. and M.Sc. degrees in computer engineering from Al-Nahrain University, Baghdad, Iraq, in 2003 and 2009 respectively.

He is currently working toward the Ph.D. in Electrical Engineering at the University of Leeds, Leeds, U.K. From (2010 – 2015) he was an assistant lecturer in Al-Mansour University College, Baghdad, Iraq and, prior to that (2007 – 2010), he was an Intelligent Network (IN), Short Message System (SMS), and (Public Switched Telephone Network) PSTN engineer with ZTE Corporation for Telecommunication, Iraq. His research interests include big data analytics, network design and energy efficiency in networks.

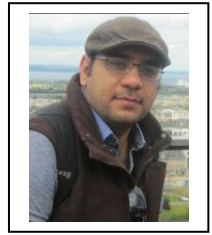

**Ahmed Q. Lawey** received the BS degree (first-class Honors) in computer engineering from the University of Al-Nahrain, Iraq, in 2002, the MSc degree (with distinction) in computer engineering from University of Al-Nahrain, Iraq, in 2005, and the PhD degree in communication networks from the University of Leeds, UK, in 2015.

From 2005 to 2010 he was a core network engineer in ZTE Corporation for Telecommunication, Iraq branch. He is currently a lecturer in communication networks in the School of Electronic and Electrical Engineer, University of Leeds. His current research interests include energy efficiency in optical and wireless networks, big data, cloud computing and Internet of Things.

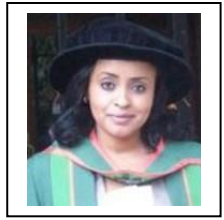

**Taisir E. H. El-Gorashi** received the B.S. degree (first-class Hons.) in electrical and electronic engineering from the University of Khartoum, Khartoum, Sudan, in 2004, the M.Sc. degree (with distinction) in photonic and communication systems from the University of Wales, Swansea, UK, in 2005, and the PhD degree in optical networking from the University of Leeds, Leeds, UK, in 2010. She is currently a Lecturer in optical networks in the School of Electrical and Electronic Engineering, University of Leeds. Previously, she held a Postdoctoral Research post at the University of Leeds (2010–2014), where she focused on the energy efficiency of optical networks investigating the use of renewable energy in core networks, green IP over WDM networks with datacenters, energy efficient physical topology design, energy efficiency of content distribution networks, distributed cloud computing, network virtualization and Big Data. In 2012, she was a BT Research Fellow, where she developed energy efficient hybrid wireless-optical broadband access networks and explored the dynamics of TV viewing behavior and program popularity. The energy efficiency techniques developed during her postdoctoral research contributed 3 out of the 8 carefully chosen core network energy efficiency improvement measures recommended by the GreenTouch consortium for every operator network worldwide. Her work led to several invited talks at GreenTouch, Bell Labs, Optical Network Design and Modelling conference, Optical Fiber Communications conference, International Conference on Computer Communications and EU Future Internet Assembly and collaboration with Alcatel Lucent and Huawei.

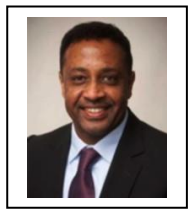

**Jaafar M. H. Elmirghani** (M' 92–SM' 99) is the Director of the Institute of Communication and Power Networks within the School of Electronic and Electrical Engineering, University of Leeds, UK. He joined Leeds in 2007 and prior to that (2000–2007) as chair in optical communications at the University of Wales Swansea he founded, developed and directed the Institute of Advanced Telecommunications and the Technium Digital (TD), a technology incubator/spin-off hub. He has provided outstanding leadership in a number of large research projects at the IAT and TD.

He received the BSc in Electrical Engineering, First Class Honours from the University of Khartoum in 1989 and was awarded all 4 prizes in the department for academic distinction. He received the PhD in the synchronization of optical systems and optical receiver design from the University of Huddersfield UK in 1994 and the DSc in Communication Systems and Networks from University of Leeds, UK, in 2014. He has co-authored Photonic switching Technology: Systems and Networks, (Wiley) and has published over 450 papers. He has research interests in optical systems and networks.

Prof. Elmirghani is Fellow of the IET, Chartered Engineer, Fellow of the Institute of Physics and Senior Member of IEEE. He was Chairman of IEEE Comsoc Transmission Access and Optical Systems technical committee and was Chairman of IEEE Comsoc Signal Processing and Communications Electronics technical committee, and an editor of IEEE Communications Magazine. He was founding Chair of the Advanced Signal Processing for Communication Symposium which started at IEEE GLOBECOM'99 and has continued since at every ICC and GLOBECOM. Prof. Elmirghani was also founding Chair of the first IEEE ICC/GLOBECOM optical symposium at GLOBECOM'00, the Future Photonic Network Technologies, Architectures and Protocols Symposium. He chaired this Symposium, which continues to date under different names. He was the founding chair of the first Green Track at ICC/GLOBECOM at GLOBECOM 2011, and is Chair of the IEEE Green ICT initiative within the IEEE Technical Activities Board (TAB) Future Directions Committee (FDC), a pan IEEE Societies initiative responsible for Green ICT activities across IEEE, 2012-present. He is and has been on the technical program committee of 34 IEEE ICC/GLOBECOM conferences between 1995 and 2016 including 15 times as Symposium Chair. He has given over 55 invited and keynote talks over the past 8 years.

He received the IEEE Communications Society Hal Sobol award, the IEEE Comsoc Chapter Achievement award for excellence in chapter activities (both in international competition in 2005), the University of Wales Swansea Outstanding Research Achievement Award, 2006; and received in international competition: the IEEE Communications Society Signal Processing and Communication Electronics outstanding service award, 2009, a best paper award at IEEE ICC'2013. Related to Green Communications he received (i) the IEEE Comsoc Transmission Access and Optical Systems outstanding Service award 2015 in recognition of "Leadership and Contributions to the Area of Green Communications", (ii) the GreenTouch 1000x award in 2015 for "pioneering research contributions to the field of energy efficiency in telecommunications", (iii) the IET 2016 Premium Award for best paper in IET Optoelectronics and (iv) shared the 2016 Edison Award in the collective disruption category with a team of 6 from GreenTouch for their joint work on the GreenMeter.

He is currently an editor of: IET Optoelectronics and Journal of Optical Communications, and was editor of IEEE Communications Surveys and Tutorials and IEEE Journal on Selected Areas in Communications series on Green Communications and Networking. He was Co-Chair of the GreenTouch Wired, Core and Access Networks Working Group, an adviser to the Commonwealth Scholarship Commission, member of the Royal Society International Joint Projects Panel and member of the Engineering and Physical Sciences Research Council (EPSRC) College. He has been awarded in excess of £22 million in grants to date from EPSRC, the EU and industry and has held prestigious fellowships funded by the Royal Society and by BT. He was an IEEE Comsoc Distinguished Lecturer 2013-2016.